\documentclass[12pt]{article}

\usepackage{graphics}
\usepackage{epsfig}
\usepackage{amssymb,amsmath}
\usepackage{array}
\usepackage{url}

\newcommand{\bea}{\begin{eqnarray}}
\newcommand{\eea}{\end{eqnarray}}

 \makeatletter
\def\alt{\mathrel{\mathpalette\gl@align<}}
\def\agt{\mathrel{\mathpalette\gl@align>}}
\def\gl@align#1#2{\lower.6ex\vbox{\baselineskip\z@skip\lineskip\z@
\ialign{$\m@th#1\hfil##\hfil$\crcr#2\crcr\sim\crcr}}} \makeatother

\begin{document}

%

%
\vspace*{1.0cm}
\begin{center}
\baselineskip 20pt

{\Large\bf Color Triplet Diquarks at the LHC}

\vspace{1cm}

{\large
Ilia Gogoladze$^{a}$\footnote{ On  leave of absence from:
Andronikashvili Institute of Physics,  Tbilisi, Georgia.},
Yukihiro Mimura$^{a}$,
%
Nobuchika Okada$^{b}$
%
and
Qaisar Shafi$^{a}$}

\vspace{1.0cm}

{
\it
$^a$Bartol Research Institute, Department of Physics and Astronomy, \\
University of Delaware, Newark, DE 19716, USA \\ \vspace{2mm}
$^b$Department of Physics and Astronomy, University of Alabama, Tuscaloosa, AL 35487, USA
}

\vspace{.5cm}

\vspace{1.5cm} {\bf Abstract}
\end{center}

\noindent
We consider a class of supersymmetric models containing
baryon number violating processes such as observable $n$-$\bar n$
oscillations that are mediated by color triplet diquark fields.
For plausible values of the diquark-quark couplings, the scalar diquark
with mass between a few hundred GeV and one TeV or so can be produced
in the $s$-channel at the LHC and detected through its decay into
a top  quark and  a hadronic jet.

\thispagestyle{empty}
\newpage


\baselineskip 18pt

\section{Introduction}

Being a proton-proton ($pp$) collider
the LHC is an ideal hunting ground for new elementary
particles which carry non-zero baryon number.
Such particles often arise as vectorlike pairs in a large class of
supersymmetric (SUSY) models,
  including some which are obtained from compactification of
superstring theories \cite{witten, Angelopoulos:1986uq, Babu:1996zv}.

The TeV scale vectorlike particles have various indirect
phenomenological applications.
They can help to substantially increase the mass
of the lightest CP-even Higgs boson of minimal SUSY standard model
(MSSM)~\cite{Babu:2008ge,Moroi:1992zk}.
The new Higgs coupling to these particles can also  modify
 the first order electroweak phase transition,
and thereby make the scalar top quark mass bound milder
in the electroweak baryogenesis scenario \cite{Ham:2010tr}.
The leptonic vectorlike particles can
generate sizable contribution to the tau and tau neutrino
electric dipole moment which can be tested in future experiment \cite{Ibrahim:2010va}.

The low scale vectorlike particle can also induce observable neutron-antineutron ($n$-$\bar n$) oscillations \cite{Chacko:1998td,Ajaib:2009fq},
in which the so-called diquark couplings play a crucial role.
The vectorlike field which couples
to two anti-quarks have baryon number 2/3
(namely twice the baryon number of quark),
and thus we refer to such a colored vectorlike field as ``diquark".
The coupling between diquarks and a pair of quarks
can make the creation and decay modes of the diquarks
very interesting in view of collider physics \cite{DelNobile:2009st}.
If the diquark fields are pair produced at the LHC
through QCD processes \cite{Chen:2008hh,Tanaka:1991nr},
they decay into quarks via
the diquark interaction.
This interaction also plays a key role in the resonant production
of the diquark field from $pp$ collision,
and if the cross section 
 is large enough to distinguish  the peak
from the standard model (SM) background, one can detect the diquark field
and determine its mass \cite{Mohapatra:2007af}.

There are two possibilities for the representation of the diquark fields:
(1) color triplet, and (2) color sextet.
The color sextet diquark fields can be embedded
in the Higgs multiplet required to break the gauge symmetry  $SU(4)_c \times SU(2)_L \times SU(2)_R$
\cite{Pati:1974yy}
down to SM,
and the mass of the diquark can remain around a TeV scale
if there exists an accidental global symmetry in the
Higgs potential \cite{Chacko:1998td}.
Such sextet diquarks do not have a renormalizable coupling
with leptons, and their existence at low energy is not problematic.
The color triplet diquarks are also well motivated
since they can be accommodated  into matter representations
in unified models~\cite{Angelopoulos:1986uq}.
However,
they can couple to both quarks and leptons,
and the most general couplings can cause  serious problems with
 rapid nucleon decay if  the diquark mass is around the TeV scale.
This difficulty is remedied by employing a $U(1)$ symmetry to prohibit the couplings of the color triplets
with leptons~\cite{Ajaib:2009fq}.
In this case,
there are no dangerous  nucleon decay operators,
but the $U(1)$ symmetry can allow  baryon number violating
processes~\cite{Ajaib:2009fq} like $n$-$\bar n$ oscillations.

In this paper, motivated by the physics of $n$-$\bar n$ oscillations,
 we study the resonant production and
 decay of the triplet diquarks at the LHC.
The attractive feature of gauge coupling unification in
MSSM is easily maintained  \cite{Babu:1996zv}
in the case of triplet diquarks accompanied by electroweak  doublets.
Clearly, the
diquark couplings with the up and down quarks
should be large enough to detect its resonant production.
However, some of the couplings (especially for the first and second generations)
are constrained by the experimental data
on meson-antimeson ($K$-$\bar K$, $B$-$\bar B$ and $D^0$-$\bar D^0$)
mixings,
and thus we will first discuss the experimentally allowed parameter range
of the diquark couplings.
In the case of sextet diquark, $D^0$-$\bar D^0$ mixing is generated
even at tree-level, while for the triplet diquarks,
the mixings are generated by box or penguin diagrams, quite similar
to the conventional analysis in MSSM.
We will study whether one can see the resonant peak from the SM background
in the phenomenologically allowed region of the diquark coupling.
If the triplet diquarks couple with quarks for both first and third generations,
one can observe resonant decay events for
both hadronic dijet and single top quark plus one jet.
Note that the triplet diquark can couple to both left- and
right-handed quarks, while the sextet diquarks can couple only with the
right-handed quarks.
It is therefore  important to observe top quark helicity to distinguish
these  two cases of diquark production from each other.

This paper is organized as follows:
In section 2,
we introduce following \cite{Ajaib:2009fq} a $U(1)$ symmetry which forbids
rapid nucleon decay processes
but allows  $n$-$\bar n$ oscillations.
In section 3,
the phenomenological constraints of the diquark couplings
are considered.
The diquark can couple with the  quarks of all generations.
In section 4,
we study the collider phenomenology of the triplet diquarks.
We first study the mass bounds on  diquarks arising  from the Tevatron results,
and then discuss  the resonant production and decays of the diquarks
at the LHC.
Section 5 summarizes our conclusions.

\section{A model with triplet diquark couplings}

We introduce the following vectorlike fields (more precisely chiral superfield) with masses in the few hundred GeV to TeV range:
\begin{equation}
D : \left({\bf 3},{\bf 1}\right)_{-\frac13} +
\bar D : \left(\bar{\bf 3}, {\bf 1}\right)_{\frac13} +
L : \left({\bf 1},{\bf 2}\right)_{-\frac12}  +
\bar L : \left({\bf 1}, {\bf 2}\right)_{\frac12}.
\end{equation}
Because $\bar D+ L$ has the same particle content
as the {\bf 5}-dimensional $SU(5)$ multiplet,
 gauge coupling unification in MSSM is preserved.
The vectorlike fields can couple with the quark and lepton fields as follows:
\begin{equation}
D q q + D u^c e^c + \bar D q \ell + \bar D u^c d^c +
L \ell e^c + L q d^c + \bar L q u^c.
\label{vector-q-q}
\end{equation}
Here we use the standard notation ($q,u^c,d^c,\ell,e^c$) for the quarks and leptons.
In general, the  couplings  in Eq.(\ref{vector-q-q})
 contradict the current experimental data~\cite{Nishino:2009gd}
unless the coupling coefficients  are less than $O(10^{-13})$.
Indeed, the  exchange of the scalar component of $D$ ($\bar D$) chiral multiplet
can generate the dimension-six proton decay operator
$q q u^{c\dagger} e^{c\dagger}$ ($q\ell u^{c\dagger} d^{c\dagger}$),
which  leads  to an  unacceptable proton decay rate.

The low scale vectorlike multiplets (${\bf 5}+\bar{\bf 5}$) can,
in general, generate neutrino masses
through loop effects if the following interactions are allowed:
\begin{equation}
 q\bar D H_d + L e^c H_d + \bar D q \ell + L \ell e^c,
\label{vector-q-2}
\end{equation}
where $H_d$ stands for the MSSM down-type Higgs field.
In this case,
one has to assume the couplings in Eq.(\ref{vector-q-2}) to be $O(10^{-4})$
to obtain the correct neutrino masses.
Since our goal in this paper is to investigate resonant production of diquark at the LHC,
we desire to have at least $O(0.1)$ strength couplings of diquarks with the SM quarks.

Thus, it seems natural to forbid rapid  proton decay and large neutrino masses
by a symmetry.
Clearly, if odd $R$-parity is assigned to the vectorlike fields,
the couplings in Eq.(\ref{vector-q-q}) are all forbidden,
which clearly is not our goal.
It was shown in Ref.~\cite{Ajaib:2009fq} that
a $(nB+mL)/2$  $U(1)$ symmetry, where $B$ and $L$ are the baryon and
lepton numbers of the fields and $n$ and $m$ are integers,
 can allow us to retain   the ``diquark couplings",
$D qq$ and $\bar D u^c d^c$,
but forbid the ``leptoquark couplings", $D u^c e^c$ and $\bar D q\ell$.
As a consequence, the rapid nucleon decay operators are forbidden.
If odd $R$-parities are assigned
for $L+\bar L$ and even $R$-parity for $D+\bar D$,
 then all the terms in Eq.(\ref{vector-q-2})
can be forbidden and therefore neutrino masses will not be  generated
from the loop effects.

Let us describe in more detail how the $(nB+mL)/2$ symmetry works.
 The charge assignments for quarks, leptons, Higgs and vectorlike fields
under $(nB+mL)/2$ symmetry are given in Table \ref{Table-1}.
\begin{table}[tbp]
\center
	\begin{tabular}{|c|c|c|c|c|c|c|c|c|c|c|c|c|} \hline
	 $q$ & $u^c$ & $d^c$ & $\ell$ & $e^c$ & $\nu^c$
         & $H_u$ & $H_d$ & $D$ & $\bar D$ & $L$ & $\bar L$ & $S$ \\ \hline
	 $\frac{n}{6}$ & $-\frac{n}{6}$ & $-\frac{n}{6}$ &
	 $\frac{m}{2}$ & $-\frac{m}{2}$ & $-\frac{m}{2}$ & 0 & 0 &
	   $-\frac{n}{3}- \frac{X_D}{2}$ & $\frac{n}{3}+ \frac{X_D}{2}$ &
	   $-\frac{n}{2}- \frac{X_L}{2}$ & $\frac{n}{2}+ \frac{X_L}{2}$ & 1 \\ \hline
	\end{tabular}
\caption{Particle charge assignments under $(nB+mL)/2$ symmetry.
The numbers $n$, $m$, $X_D$ and $X_L$ are all integers.}
\label{Table-1}
\end{table}
The $U(1)$ symmetry is anomalous and the Green-Schwarz mechanism \cite{GS} is applied.
The SM singlet field $S$ acquires a vacuum expectation value (VEV)
close to the string scale $M_{st}$
from the
Fayet-Illiopoulos $D$-term potential.
If $n+m$ is an odd integer,  the $\Delta B=\Delta L = \pm 1$ operators (e.g.,
$qqq \ell$, $u^c d^c u^c e^c$, $q q u^{c\dagger} e^{c\dagger}$)
are all forbidden,
as long as no field with fractional charge of $(nB+mL)/2$ symmetry
acquires a VEV.
The couplings of quarks and  leptons with $D+\bar D$ fields are expressed as
\begin{equation}
\left(\frac{S}{M_{st}}\right)^{\frac{X_D}{2}} D q q+
\left(\frac{S}{M_{st}}\right)^{\frac{X_D}{2}} \bar D u^c d^c
+\left(\frac{S}{M_{st}}\right)^{\frac{X_D+n+m}{2}} D u^c e^c
+\left(\frac{S}{M_{st}}\right)^{-\frac{X_D+n+m}{2}} \bar D q \ell.
\end{equation}
If the exponents of $S$ are not integers, the respective
couplings are not allowed.
As one can see, the diquark couplings are allowed only
if $n+m$ is an odd number and $X_D$ is even.

It is interesting to note that
the $(nB+mL)/2$ symmetry forbids rapid nucleon decay,
but it can allow  baryon number violating processes
like the $\Delta B=2$ $n$-$\bar n$ oscillations.
In fact, the appropriate operators responsible for  $n$-$\bar n$ oscillations
are effectively generated \cite{Ajaib:2009fq}
if there are MSSM singlet fields $N:({\bf 1},{\bf 1})_0$,
in addition to the diquark fields.
The typical $n$-$\bar n$ oscillation operators
are $u^c d^c d^c u^c d^c d^c$ and
$q q d^{c\dagger} q q d^{c\dagger}$,
and the diquark couplings, $\bar D u^c d^c$ and $Dqq$,
 play a crucial role in generating  them at  tree level.
Through such couplings, the diquark can decay
 into pairs of up-type and down-type quarks
 and generate a hadronic dijet or a single top quark
 plus a hadronic jet.
%

In order to induce the $n$-$\bar n$ oscillation operators,
one can also add a pair of vectorlike fields
$\bar U:(\bar{\bf3}, {\bf1})_{-2/3}+U:({\bf3},{\bf 1})_{2/3}$,
and a diquark coupling $\bar U d^c_i d^c_j$.
Since the representations are included in $\bf 10+\overline{\bf 10}$
 multiplets of $SU(5)$ symmetry,
the MSSM gauge coupling unification can be maintained if
the complete contents of $\bf 10+\overline{\bf 10}$ representations are introduced.
Through the diquark coupling,
the $\bar U$ diquark decay produces only hadronic dijet.
Although at high energy hadron colliders
 the dijet signal from the resonant production of
 the diquarks seems to be overwhelmed by QCD background,
 the large luminosity at the Tevatron  allows us
 to give a fairly  severe lower bound on the  diquark mass
 \cite{Aaltonen:2008dn}.
Depending on the couplings,
 the diquarks can be discovered at the LHC
 through the dijet events \cite{Atag:1998xq}.
Noting that the $D+\bar D$ diquarks can decay
 into single top quark plus a hadronic jet,
 we concentrate in this paper on collider phenomenology
 of these diquarks and their decays into top quarks.
We find  that this process involving the  top quark
 gives a lower bound on the diquark mass which is
 more severe than that obtained from the dijet process.
Furthermore, the final state top quarks are useful
 in identifying  the baryon number of the diquark
 and a chiral structure of the coupling between
 the diquarks and quarks.

\section{Phenomenological constraint on the diquark couplings}

The magnitude of the coupling between the  diquarks and quarks
is clearly important for the production and decay modes of the diquark
fields at the LHC,
and we therefore first study the phenomenological constraints
 on these couplings.
We define the diquark coupling constants as
\begin{equation}
\frac{1}{2} \kappa_{ij} D q_i q_j +  \bar \kappa_{ij} \bar D u^c_i d^c_j,
\end{equation}
where $i,j$ are flavor indices.
The most important constraint on the coupling constants
and the vectorlike mass of $D,\bar D$ arise
from the oblique corrections for the
weak gauge bosons ($S$ and $T$ parameters) \cite{Lavoura:1992np,PDG}.
The mass and couplings of the diquarks
 are also constrained from the Tevatron experiment,
which will be discussed in the next section.

In SUSY models,
the diquark couplings can induce
flavor non-universality in SUSY breaking squark mass matrices
through the renormalization group (RG) evolutions,
even if universality is assumed at the unification scale.
The flavor violation effects can generate processes
via the flavor changing neutral current (FCNC)
(e.g., meson-antimeson mixings and $b\to s(d) \gamma$)
through loop diagrams involving  SUSY particles.
Therefore, generic $O(0.1)$ coupling coefficients  for the first and second generations
are disfavored.
In addition to such indirect effects of FCNC,
the diquark couplings can generate FCNC loop diagrams
in which the diquark fields propagate.

We note that $D^0$-$\bar D^0$ meson mixing is generated even at the tree level
if a sextet diquark field $\Delta : ({\bf 6},{\bf 1})_{\frac43}$
and the coupling $\Delta u^c_i u^c_j$ are considered.
However, in the case of triplet diquarks (including $U+\bar U$),
all FCNC processes are generated at the loop level,
and the essential features for finding the experimental bounds are
 similar to the  MSSM case \cite{Gabbiani:1988rb}.
Thus the matrix elements of $\kappa \kappa^\dagger$, $\kappa \bar \kappa$,
etc.,
are constrained for a given mass spectrum.

If the diquark coupling matrix elements are hierarchical,
similar to the Yukawa couplings for quarks and leptons,
the FCNC effects can be adequately suppressed.
However, 
for the resonant production of diquarks at the LHC,
we prefer a parameter region where
the couplings for the first generation
$\kappa_{11}$ (or $\kappa_{12}$) are of $O(0.1)$.
We make the following simplifying  ansatz for
 minimal flavor violation:
\begin{equation}
\kappa_{ij} = \kappa \delta_{ij}, \qquad
\bar \kappa_{ij} = \bar\kappa \delta_{ij}.
\label{assumption}
\end{equation}
%
In this case,
the RG evolutions do not generate any new sources
of FCNC in the squark mass matrices.
The new direct loop contributions via the diquark fields,
however, are generated from the left-handed coupling
$\kappa D qq$
because of the Cabibbo-Kobayashi-Maskawa (CKM) mixings.
The contributions to  FCNC processes under the
assumption in Eq.(\ref{assumption}) is similar to the chargino
loop diagram (except for the charge and couplings),
and thus the amount of the contribution is not significant
compared to that of the SM
when  flavor universality of the squark squared masses
is assumed at a unification scale.
Therefore, for minimal flavor violation,
the experimental constraint is satisfied
as long as
the diquarks are heavier than about 300 GeV,
and $\kappa 
\leq 1$.
Actually, if the diquark coupling is more than $O(0.1)$,
the diquark should be heavier than about 300 GeV
from the Tevatron result,
as we will see in next section.

Any deviation from Eq.(\ref{assumption}) is severely constrained
from  $K$-$\bar K$ mixings,
especially for $\kappa_{11}- \kappa_{22}$, $\kappa_{12}$
and $\bar \kappa_{12}$ $\alt O(10^{-3})-O(10^{-4})$.
The $B_{d,s}$-$\bar B_{d,s}$ mixings and $b\to s(d) \gamma$ decay
give $\kappa_{23}$, $\bar \kappa_{23}$ $\alt O(10^{-2})$,
and $\kappa_{13}$, $\bar \kappa_{13}$ $\alt O(10^{-3})$.
Such small contributions in the flavor off-diagonal elements
are not crucial for the collider studies, and
so will be disregarded in what follows.

Through the RG running,
the squark masses become smaller at low energy, for a given boundary condition,
as  the diquark couplings get larger.
Under the assumption in
Eq.(\ref{assumption}),
the SUSY breaking squared scalar masses of diquarks
become negative when the coupling is $O(1)$ at low energy
due to the fact that the wave function renormalizations
of the diquark fields are proportional to Tr$\,(\kappa \kappa^\dagger)$
and Tr$\,(\bar\kappa \bar\kappa^\dagger)$.
To avoid color and charge symmetry breaking,
an appropriate size  vectorlike mass of
the diquark them  needs to be added,
in which case the  fermionic partners of the diquarks are heavier than
the scalar diquarks.
If the SUSY breaking diquark squared masses are negative,
 the squarks do not become tachyonic even if the diquark coupling
is large.
If the diquark couplings are about 0.3 at low energy,
the squared scalar mass will not become negative due to the gluino loop contribution
and the fermionic diquark is lighter than the scalar diquark.

\section{Collider phenomenology}

Since the scalar diquarks ($D$ and $\bar{D}$) couple
 with a pair of quarks, they can be produced
 in the $s$-channel at the Tevatron and the LHC
 through  annihilation of a pair of quarks.
As a signature of scalar diquark productions at hadron colliders,
 we concentrate on its decay channel which includes
 a single top or anti-top quark in the final state.
Because of its  mass, the top quark decays electroweakly
 before hadronizing.
Due to this characteristic feature not shared by the  other quarks, the top quark can be 
 an ideal tool \cite{TopPhys} to probe new physics
 beyond the standard model \cite{tp}.

We consider here the scalar diquark ($D$) production and
 the analysis for $\bar D$ is basically the same,
 except for the chirality of the initial and final quarks.
The fundamental processes 
 are
 $u d \to D \to tb$
($\bar{u} \bar{d} \to D^\dagger \to \bar{t} \bar{b}$
 for the scalar anti-diquark production).
The cross section at the parton level is given by
\bea
 \frac{ d \sigma( u d \rightarrow D \rightarrow t b)}
 {d \cos \theta}
 = \frac{|\kappa|^4}{48 \pi}
   \frac{(\hat{s}- m_t^2)^2}
    {(\hat{s}-M_D^2)^2
  + M_D^2 \Gamma_{\rm tot}^2 } .
\label{CrossParton}
\eea
Here, we have neglected all quark masses except
 for top quark mass ($m_t=173.1$ GeV \cite{topmass}),
 $\theta$ is the scattering angle,
%
 and the total decay width $\Gamma_{\rm tot}$ of
 the scalar diquark is the sum of its partial decay widths,
\bea
 \Gamma (D \to u d, c s)
 &=& \frac{1}{8 \pi}  |\kappa|^2 \; M_D,   \nonumber \\
 \Gamma (D \to t b)
 &=& \frac{1}{8 \pi}  |\kappa|^2 \; M_D
   \left(  1- \frac{m_t^2}{M_D^2} \right)^2.
\eea
Note that the cross section is independent of the scattering angle
 because the diquark is a scalar.

At the  Tevatron, the total production cross section
 of a quark pair ($t b$) through the scalar diquark
 in the $s$-channel is given by
\bea
 \sigma (p \bar{p} \to t b + X)
 &=& \int dx_1 \int dx_2 \int d \cos \theta
 \left[
   f_u(x_1, Q^2)  f_{\bar d}(x_2, Q^2)
 + f_{\bar u}(x_1, Q^2)  f_d(x_2, Q^2)
 \right]  \nonumber \\
 &\times &
  \frac{d \sigma(u d \to D \to t b;
  \hat{s}=x_1 x_2 E_{\rm CMS}^2)}{d \cos \theta},
\label{CrossTevatron}
\eea
where $f_q$ denotes the parton distribution function
 for a quark $q$ in a proton,
 and $E_{\rm CMS}$ is the collider energy.
Note that one parton distribution function is for the up (down) quark
 and the other is for the sea down (up) quark.
At the Tevatron the production cross
 sections of
 $D$ and $D^\dagger$ are the same,
 reflecting the fact that the total baryon number of the
 initial state  $p \bar{p}$ is zero.

At the  LHC, the total $tb$ production cross section
 is given by
\bea
 \sigma (p p \to t b +X)
 &=& \int dx_1 \int dx_2 \int d \cos \theta \nonumber \\
 &\times &  2 f_u(x_1, Q^2)  f_d(x_2, Q^2)
  \frac{d \sigma(u d \to D \to t b;
  \hat{s}=x_1 x_2 E_{\rm CMS}^2)}{d \cos \theta}.
\eea
Here, both parton distribution functions are
 for the valence quarks, corresponding
 to a proton-proton system at the  LHC.
The total production cross section of $\bar{t} \bar{b}$
 through the scalar anti-diquark production in the $s$-channel
 is obtained by replacing with the appropriate parton distribution function
 for anti-quarks.
The initial $p p$ state has a positive baryon number,
 so that the production cross section at the LHC of $D$
 is much larger than the one for $D^\dagger$.
The dependence of the cross section on
 the final state invariant mass $M_{\rm inv}$
 is given by 
\bea
 \frac{d \sigma (p p \to t b + X)}{d M_{\rm inv}}
= \int^1_{ \frac{M_{\rm inv}^2}{E_{\rm CMS}^2}} dx
 \frac{4 M_{\rm inv}}{x E_{\rm CMS}^2}
  f_u(x, Q^2) f_d
 \left( \frac{M_{\rm inv}^2}{x E_{\rm CMS}^2}, Q^2
 \right)
 \sigma(u d \to D \to t b; \hat{s}=M_{\rm inv}^2 ).
\label{CrossLHC}
\eea

Let us first examine the lower bound on the scalar diquark mass
 from the Tevatron.
The production of single top quark at the Tevatron at
 $\sqrt{s} =1.96$ TeV has been observed,
 and the measured cross section is 
 consistent with the standard model prediction \cite{SingleTop}.
We consider the cross section \cite{D0X},
\bea
 \sigma(p {\bar p} \to t b + X) =1.05 \pm 0.81 \; {\rm pb},
\eea
 from $s$-channel W boson, and 
 use it as a constraint
 on $tb$ production cross section
 through the scalar diquark in the $s$-channel.
Any possible new physics
 should be in the uncertainty range of
 this cross section, and so we take the bound to be
\bea
\sigma( p \bar{p} \to D \to t b) \lesssim 0.81 {\rm pb}.
\eea
In our numerical analysis, we employ CTEQ5M \cite{CTEQ}
 for the parton distribution functions
 with the factorization scale $Q=m_t=173.1$ GeV.
Figure~1 shows the total cross section of $tb$ production
 as a function of the scalar diquark mass,
 with $E_{\rm CMS} = 1.96$ TeV.
The lower bound on the diquark mass is found to be
352, 415,  and 459 GeV, for $\kappa$ = 0.2, 0.3, 0.4, respectively.
For these values of $\kappa$, we also evaluate
 the dijet production cross section via the resonant
 diquark production and find that the signal cross section
 is below the observed upper limit \cite{Aaltonen:2008dn}
 and, hence, there  no lower bound is obtained.
We also find that for a sufficiently large value of $\kappa$,
 the lower bound on the diquark mass obtained by
 the observed cross section of single top quark production
 is always more severe than that from
 the dijet production cross section.

\begin{figure}[t]
\center
\includegraphics[width=0.6 \columnwidth=0.4]{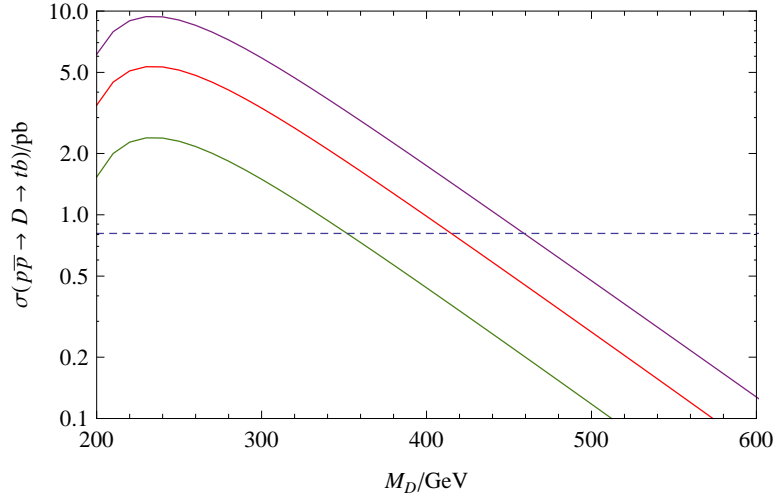}
\caption{
The cross section of $tb$ production at the Tevatron
 with $E_{\rm CMS}=1.96$ TeV mediated
 by the scalar diquark in the $s$-channel.
Solid lines correspond to  $\kappa$ = 0.2, 0.3, 0.4
from bottom to top.
The horizontal dotted line shows the Tevatron bound.
}
\label{Fig1}
\end{figure}

\begin{figure}[t]
\center
\includegraphics[width=0.68 \columnwidth,viewport = 0 0 315 184]{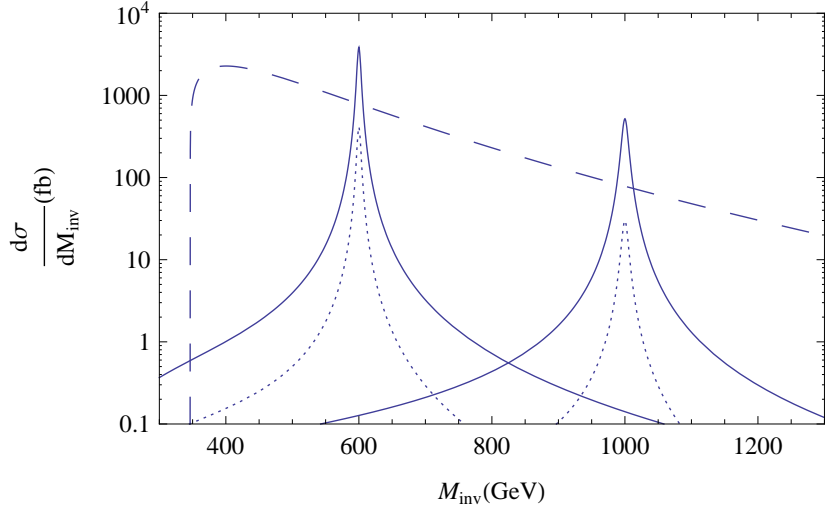}
\caption{
 The differential cross sections for $tb$ (solid line),
 $\bar{t} \bar{b}$ (dotted line)
 versus the invariant mass of the final states.
The left peak corresponds to $M_D=600$ GeV
 and the right one to $M_D=1$ TeV.
The dashed line is the standard model $t\overline{t}$ background. Here $\kappa=0.3$
}
\label{Fig2}
\end{figure}

Next we investigate the scalar diquark and anti-diquark productions
 at the LHC with $E_{\rm CMS}=14$ TeV.
The differential cross sections for each process
 for $\kappa=0.3$ and  with $M_D=600$ GeV and 1 TeV are depicted in Figure 2.
Here, we compare the single top quark production
 via the scalar diquark with the $t \overline{t}$
 production cross section in the standard model.
  At the LHC, the main background for the single top quark production  is
$t\bar{t}$ production \cite{SSW}, and it is easy to misidentify $t\bar{t}$ events as single top events.
So, for a conservative analysis, we   compare our signal to
$t\bar{t}$ production. The single top production cross section is smaller than the  $t\bar{t}$ production
cross section at the LHC by factor of about 7.

We can see that the peak cross sections for $tb$ production
 exceeds the standard model cross section,
 while the $\bar{t} \bar{b}$ cross section by comparison is lower
 (however, the $\bar{t} \bar{b}$ cross section deviates sizably
 from the standard model cross section).
This discrepancy between the production cross sections
 of scalar diquark and anti-diquark at the LHC is direct evidence
 for non-zero baryon number of the scalar diquark.
Note that one can distinguish the top quark from the anti-top quark
 through their semi-leptonic decays.
Counting the number of top quark events and anti-top quark events
 would reveal a non-zero baryon number for the scalar diquark.

The result for the production of the scalar diquark $\bar D$
 is the same when we take $\kappa=\bar{\kappa}$
 and replace $D^\dagger$ with $\bar D$.
However, there is a crucial difference
 in the spin polarization of the final state top (anti-top) quark.
This is  because while $D$ couples with the left-handed top quark,
 ${\bar D}^\dagger$ couples with the right-handed top quark.
Since the top quark decays before hadronizing,
 the information of the top quark spin polarization is
 directly transferred to its decay products and
 results in significant angular correlations
 between the top quark polarization axis and
 the direction of motion of the decay products~\cite{TopSpin}.
It has been shown that measuring the top spin correlations
 can increase the sensitivity for detecting a new particle
 at the Tevatron \cite{SpinCorr1} and the LHC~\cite{SpinCorr2}.
For the scalar diquark production,
 we can distinguish between  $D$ and $\bar D^\dagger$ productions
 by measuring the polarization of top quarks
 from their decays.
It is interesting that only the right-handed top quark
 is produced by the scalar diquark ${\bar D}^\dagger$ decay,
 while the single top quark produced 
 through the standard model process is purely left-handed.

We have discussed the scalar diquark production
 in the $s$-channel, where the production cross section
 is controlled by the diquark coupling.
Because they carry color, diquarks can be
 pair produced through QCD processes,
 and this production is independent of the diquark coupling.
For the sextet diquark, the pair production at the LHC
 has been investigated in detail \cite{Chen:2008hh}.
Although the analysis there focused on $t t \bar{t} \bar{t}$
 final state, a similar analysis here for
  the pair production of triplet scalar diquarks is applicable.
The total pair production cross section of the triplet diquark
 is about an order of magnitude smaller than
 the sextet diquark case because of the difference
 of the $SU(3)$ group factors (see Fig.~2 in \cite{Chen:2008hh}).

If kinematically allowed, the scalar diquark  can also decay
 into a pair of squarks through the soft SUSY breaking
 trilinear coupling.
In this case, the scalar diquark resonance can enhance
 the production cross section of squarks
 and can give some advantage to sparticle searches at the LHC.

Fermion diquarks can be pair produced through QCD processes
 and the production cross section is of the same order of
 magnitude as the production cross section
 of the scalar diquark pair.
Since they are $R$-parity odd particles,
 the decay process of fermion diquarks is similar to the
 usual sparticle decay processes.
For example, a fermion diquark can decay into
 squarks and quarks, and this process is similar to
 gluino decay, except for the difference arising from  baryon number.
Since the fermion diquark  baryon number is  twice that  of the quark,
 it is crucial to identify its baryon number
 from its decay products.
However, this can be quite challenging.

\section{Conclusion}

We have investigated a class of SUSY models which contain color triplet diquark fields with masses of around a few hundred GeV to one TeV.
 The couplings between the diquark and  a pair of quarks
 play a crucial role in observable $n$-$\bar n$ oscillation
 which may be tested in the near future.
Based on a relatively simple model recently proposed in \cite{Ajaib:2009fq},
 we have investigated the collider phenomenology of the  diquark fields.
The phenomenological constraints require that the couplings
 between the diquarks and a pair of quarks should be
 almost flavor-blind and their magnitude should be less than unity.
For the diquark couplings satisfying these constraints,
 we have analyzed $s$-channel scalar diquark production
 and its subsequent decay into the top and bottom quarks
 at hadron colliders.
We first investigated the lower bound on the scalar diquark mass
 from the current observation of single top quark production
 at the Tevatron. We find that 
 the diquark mass $M_D \agt 415$ GeV
 for a diquark coupling constant of $0.3$.
At the LHC, the differential cross section of the scalar diquark
 production shows a resonance peak 
 for a diquark coupling of ${\cal O}(0.1)$.
The difference in cross sections for
 scalar diquark and anti-diquark at the LHC,
 which can be identified by comparing the number of
 events with top and antitop quarks, provides direct evidence
 of nonzero baryon number of the diquark.
The scalar diquarks $D$ and ${\bar D}^\dagger$
 are identical except for the chirality of the quarks to which they couple.
Measuring the spin polarization of top quarks
 offers a way to distinguish between %
 these diquarks at the LHC.

\section*{Acknowledgments}

This work
is supported in part by the DOE Grant No. DE-FG02-91ER40626
(I.G., Y.M. and Q.S.) and GNSF Grant No. 07\_462\_4-270 (I.G.).
N.O. would like to thank the Particle Theory Group
 of the University of Delaware for hospitality
 during his visit.


\begin{thebibliography}{99}


\bibitem{witten}
See, for instance,
%
  E.~Witten,
  Nucl.\ Phys.\  B {\bf 258}, 75 (1985);
%
  G.~Lazarides, C.~Panagiotakopoulos and Q.~Shafi,
  Phys.\ Rev.\ Lett.\  {\bf 56}, 432 (1986);
%
  G.~Lazarides, P.~K.~Mohapatra, C.~Panagiotakopoulos and Q.~Shafi,
  Nucl.\ Phys.\  B {\bf 323}, 614 (1989).



\bibitem{Angelopoulos:1986uq}
 V.~D.~Angelopoulos, J.~R.~Ellis, H.~Kowalski, D.~V.~Nanopoulos,
N.~D.~Tracas and F.~Zwirner,
 Nucl.\ Phys.\  B {\bf 292}, 59 (1987);
%
  J.~L.~Hewett and T.~G.~Rizzo,
  Phys.\ Rept.\  {\bf 183}, 193 (1989);
%
  S.~F.~King, S.~Moretti and R.~Nevzorov,
  Phys.\ Rev.\  D {\bf 73}, 035009 (2006)
  [arXiv:hep-ph/0510419];
%
  J.~Kang, P.~Langacker and B.~D.~Nelson,
  Phys.\ Rev.\  D {\bf 77}, 035003 (2008)
  [arXiv:0708.2701 [hep-ph]].






\bibitem{Babu:1996zv}
  K.~S.~Babu and J.~C.~Pati,
  Phys.\ Lett.\  B {\bf 384}, 140 (1996)
  [arXiv:hep-ph/9606215];
%
%
  M.~Bastero-Gil and B.~Brahmachari,
  Nucl.\ Phys.\  B {\bf 575}, 35 (2000)
  [arXiv:hep-ph/9907318];
%
  J.~L.~Chkareuli, I.~G.~Gogoladze and A.~B.~Kobakhidze,
  Phys.\ Rev.\ Lett.\  {\bf 80}, 912 (1998);
%
  J.~L.~Chkareuli, C.~D.~Froggatt, I.~G.~Gogoladze and A.~B.~Kobakhidze,
  Nucl.\ Phys.\  B {\bf 594}, 23 (2001)
  [arXiv:hep-ph/0003007];
%
  T.~Li, D.~V.~Nanopoulos and J.~W.~Walker,
  arXiv:0910.0860 [hep-ph].











\bibitem{Babu:2008ge}
K.~S.~Babu, I.~Gogoladze, M.~U.~Rehman and Q.~Shafi,
  Phys.\ Rev.\  D {\bf 78}, 055017 (2008)
  [arXiv:0807.3055 [hep-ph]];
%
  S.~P.~Martin,
  arXiv:0910.2732 [hep-ph];
%
  P.~W.~Graham, A.~Ismail, S.~Rajendran and P.~Saraswat,
  arXiv:0910.3020 [hep-ph];
%
  K.~S.~Babu, I.~Gogoladze and C.~Kolda,
  arXiv:hep-ph/0410085.







\bibitem{Moroi:1992zk}
  T.~Moroi and Y.~Okada,
  Phys.\ Lett.\  B {\bf 295}, 73 (1992);
%
  Mod.\ Phys.\ Lett.\  A {\bf 7}, 187 (1992).




\bibitem{Ham:2010tr}
  S.~W.~Ham, S.~a.~Shim and S.~K.~OH,
  arXiv:1001.1129 [hep-ph].

\bibitem{Ibrahim:2010va}
  T.~Ibrahim and P.~Nath,
  arXiv:1001.0231 [hep-ph].



\bibitem{Chacko:1998td}
  Z.~Chacko and R.~N.~Mohapatra,
  Phys.\ Rev.\  D {\bf 59}, 055004 (1999)
  [arXiv:hep-ph/9802388];
%
  K.~S.~Babu and R.~N.~Mohapatra,
  Phys.\ Lett.\  B {\bf 518}, 269 (2001)
  [arXiv:hep-ph/0108089];
%
  K.~S.~Babu, P.~S.~Bhupal Dev and R.~N.~Mohapatra,
  Phys.\ Rev.\  D {\bf 79}, 015017 (2009)
  [arXiv:0811.3411 [hep-ph]];
%
  B.~Dutta, Y.~Mimura and R.~N.~Mohapatra,
  Phys.\ Rev.\ Lett.\  {\bf 96}, 061801 (2006)
  [arXiv:hep-ph/0510291].



\bibitem{Ajaib:2009fq}
  M.~A.~Ajaib, I.~Gogoladze, Y.~Mimura and Q.~Shafi,
  Phys.\ Rev.\  D {\bf 80}, 125026 (2009)
  [arXiv:0910.1877 [hep-ph]].



\bibitem{DelNobile:2009st}
  E.~DelNobile, R.~Franceschini, D.~Pappadopulo and A.~Strumia,
  Nucl.\ Phys.\  B {\bf 826}, 217 (2010)
  [arXiv:0908.1567 [hep-ph]];
%
  T.~Han, I.~Lewis and T.~McElmurry,
  arXiv:0909.2666 [hep-ph];
%
  A.~Arhrib, R.~Benbrik and C.~H.~Chen,
  arXiv:0911.4875 [hep-ph].





\bibitem{Chen:2008hh}
  C.~R.~Chen, W.~Klemm, V.~Rentala and K.~Wang,
  Phys.\ Rev.\  D {\bf 79}, 054002 (2009)
  [arXiv:0811.2105 [hep-ph]].



\bibitem{Tanaka:1991nr}
  H.~Tanaka and I.~Watanabe,
  Int.\ J.\ Mod.\ Phys.\  A {\bf 7}, 2679 (1992).



\bibitem{Mohapatra:2007af}
  R.~N.~Mohapatra, N.~Okada and H.~B.~Yu,
  Phys.\ Rev.\  D {\bf 77}, 011701 (2008)
  [arXiv:0709.1486 [hep-ph]].


\bibitem{Pati:1974yy}
  J.~C.~Pati and A.~Salam,
  Phys.\ Rev.\ D {\bf 10}, 275 (1974).






\bibitem{Nishino:2009gd}
  H.~Nishino {\it et al.}  [Super-Kamiokande Collaboration],
  Phys.\ Rev.\ Lett.\  {\bf 102}, 141801 (2009)
  [arXiv:0903.0676 [hep-ex]].







\bibitem{GS}
  M.~B.~Green and J.~H.~Schwarz,
  Phys.\ Lett.\  B {\bf 149}, 117 (1984);
%
  Nucl.\ Phys.\  B {\bf 255}, 93 (1985);
%
  M.~B.~Green, J.~H.~Schwarz and P.~C.~West,
  Nucl.\ Phys.\  B {\bf 254}, 327 (1985).



\bibitem{Aaltonen:2008dn}
 T.~Aaltonen {\it et al.}  [CDF Collaboration],
 Phys.\ Rev.\  D {\bf 79}, 112002 (2009)
 [arXiv:0812.4036 [hep-ex]].


\bibitem{Atag:1998xq}
 S.~Atag, O.~Cakir and S.~Sultansoy,
 Phys.\ Rev.\  D {\bf 59}, 015008 (1999);
%
  O.~Cakir and M.~Sahin,
  Phys.\ Rev.\  D {\bf 72}, 115011 (2005)
  [arXiv:hep-ph/0508205].



\bibitem{Lavoura:1992np}
L.~Lavoura and J.P.~Silva,
Phys.\ Rev.\ D {\bf 47}, 2046 (1993);
%
N.~Maekawa,
Phys.\ Rev.\ D {\bf 52}, 1684 (1995).


\bibitem{PDG}
  C.~Amsler {\it et al.}  [Particle Data Group],
  Phys.\ Lett.\  B {\bf 667}, 1 (2008).



\bibitem{Gabbiani:1988rb}
  F.~Gabbiani and A.~Masiero,
  Nucl.\ Phys.\ B {\bf 322}, 235 (1989);
%
  J.~Hagelin, S.~Kelley and T.~Tanaka,
  Nucl.\ Phys.\ B {\bf 415}, 293 (1994);
%
  F.~Gabbiani, 
  E.~Gabrielli, A.~Masiero and L.~Silvestrini,
  Nucl.\ Phys.\ B {\bf 477}, 321 (1996)
  [arXiv:hep-ph/9604387].


\bibitem{TopPhys}
  C.~T.~Hill and S.~J.~Parke,
  Phys.\ Rev.\  D {\bf 49}, 4454 (1994)
  [arXiv:hep-ph/9312324].



\bibitem{tp}
  K.~Agashe, A.~Belyaev, T.~Krupovnickas, G.~Perez and J.~Virzi,
  Phys.\ Rev.\  D {\bf 77}, 015003 (2008)
  [arXiv:hep-ph/0612015];
%
  B.~Lillie, L.~Randall and L.~T.~Wang,
  JHEP {\bf 0709}, 074 (2007)
  [arXiv:hep-ph/0701166];
%
  D.~Choudhury and D.~K.~Ghosh,
  Int.\ J.\ Mod.\ Phys.\  A {\bf 23}, 2579 (2008)
  [arXiv:0707.2074 [hep-ph]].




\bibitem{topmass}
  C.~Vellidis  [CDF Collaboration],
  arXiv:0910.3392 [hep-ex].


\bibitem{SingleTop}
T.~Aaltonen {\it et al.}  [CDF Collaboration],
  Phys.\ Rev.\ Lett.\  {\bf 101}, 252001 (2008)
  [arXiv:0809.2581 [hep-ex]];
V.~M.~Abazov {\it et al.}  [D0 Collaboration],
  Phys.\ Rev.\ Lett.\  {\bf 103}, 092001 (2009)
  [arXiv:0903.0850 [hep-ex]];
T.~Aaltonen {\it et al.}  [CDF Collaboration],
  Phys.\ Rev.\ Lett.\  {\bf 103}, 092002 (2009)
  [arXiv:0903.0885 [hep-ex]].


\bibitem{D0X}
 C.~E.~Gerber,
 arXiv:0909.4794 [hep-ex].


\bibitem{CTEQ}
%
  H.~L.~Lai {\it et al.}  [CTEQ Collaboration],
  Eur.\ Phys.\ J.\  C {\bf 12}, 375 (2000)
  [arXiv:hep-ph/9903282];
%





\bibitem{SSW}
T.~Stelzer, Z.~Sullivan and S.~Willenbrock,
  Phys.\ Rev.\  D {\bf 58}, 094021 (1998)
  [arXiv:hep-ph/9807340].


\bibitem{TopSpin}
 For studies at the LHC see, for example,
 F.~Hubaut, E.~Monnier, P.~Pralavorio, K.~Smolek and V.~Simak,
  Eur.\ Phys.\ J.\  C {\bf 44S2}, 13 (2005)
  [arXiv:hep-ex/0508061].

\bibitem{SpinCorr1}
 R.~M.~Harris, C.~T.~Hill and S.~J.~Parke,
 arXiv:hep-ph/9911288.

\bibitem{SpinCorr2}
 M.~Arai, N.~Okada, K.~Smolek and V.~Simak,
  Phys.\ Rev.\  D {\bf 70}, 115015 (2004)
  [arXiv:hep-ph/0409273];
  Phys.\ Rev.\  D {\bf 75}, 095008 (2007)
  [arXiv:hep-ph/0701155];
  Acta Phys.\ Polon.\  B {\bf 40}, 93 (2009)
  [arXiv:0804.3740 [hep-ph]];
M.~Arai, N.~Okada and K.~Smolek,
  Phys.\ Rev.\  D {\bf 79}, 074019 (2009)
  [arXiv:0902.0418 [hep-ph]].


\end{thebibliography}
\end{document}